
%
%
%
%
\input harvmac
%
%
%
%
\message{(modified format)}
\ifx\answ\bigans
\magnification=1200\unredoffs\baselineskip=14pt plus 2pt minus 1pt
\hsbody=6.5truein\hstitle=\hsbody\hsize=\hsbody\vsize=9.5truein
\else\hsbody=4.935truein\hstitle=\hsbody
\fullhsize=10.4truein\hsize=\hsbody
\fi
\def\tfontsize{scaled\magstep3}
 \tfontsize \font\titlerms=cmr7 \tfontsize
 \tfontsize  \tfontsize
 \tfontsize  \tfontsize
 \tfontsize  \tfontsize
 \tfontsize  \tfontsize
\def\Title#1#2{\tenpoint\hsize=\hstitle\rightline{#1}%
\vskip .5in\centerline{\titlefont #2}\footnotefont\vskip .35in}
\def\Date#1{\bigskip\bigskip\tenpoint\global\hsize=\hsbody%
\baselineskip=14pt plus 1pt minus .5pt}
\edef\BF{\bf}
\def\nwsec#1{\def\bf{\titlerms}\newsec{#1}\def\bf{\BF}}
%
%
\def\listrefs{\footatend\bigskip\medskip%
\immediate\closeout\rfile\writestoppt
\baselineskip=11.5pt\parskip=1.8pt plus .4pt
\noindent{{\titlerms References}}\medskip{\frenchspacing%
\parindent=10pt\escapechar=` \input
\jobname.refs\vfill\eject}\nonfrenchspacing}
%

\def\bar{\overline}

\def\({\left(}
\def\){\right)}
\def\[{\left[}
\def\]{\right]}

\def\VEV#1{\langle#1\rangle}
\def\th{^{\rm th}}

\def\exp{{\rm exp}}

\def\ubl#1{\{#1\}}
\def\usbl#1{\hbox{\bf({\rm $#1$})}}

 \def\R{{\bf R}}

\def\q{{\cal Q}} 
 
\def\a{\alpha} \def\b{\beta} \def\g{\gamma} 
  \def\l{\lambda} \def\p{\phi}
\def\t{\theta}   
\def\G{\Gamma}
\def\ha{\hat\alpha}

\def\forget#1{}
\def\B{\bullet}
\def\W{\circ}

\def\wB{w_{\{\B\}}}
\def\wW{w_{\{\W\}}}

\def\newt{\centerline{Isaac Newton Institute for Mathematical
Sciences,}\centerline{20 Clarkson Road, Cambridge CB3 0EH, UK}}

%
%
\def\clap#1{\hbox to 0pt{\hss #1\hss}}
\def\label#1{\lower2ex\clap{$\scriptstyle #1$}}
\def\ulabel#1{\raise3ex\clap{$\scriptstyle #1$}}
%
%
\def\tick#1.#2.{\vrule height#1pt depth#2pt}
\def\seg#1.#2.{{\hfil \tick#1.#2.}}
\newcount\n
\def\tline#1.#2.#3.{{\n=0 \loop\ifnum\n<#1 \advance\n by1 {\hfil
   \tick#2.#3.}\repeat}}
\def\tickline#1.#2.#3.#4.#5.#6.
 {\n=0 \vbox{\hrule \vskip-#3pt
   \hbox to #1in{\label{#5}\tick#3.#4.
   \loop\ifnum\n<#2 \advance\n by1 \seg#3.#4.\repeat\label{#6}}}}
%
%

%
%
\font\dynkfont=cmsy10 scaled\magstep4    \skewchar\dynkfont='60
\def\dynk{\textfont2=\dynkfont}
\def\hr#1,#2;{\dimen0=.4pt\advance\dimen0by-#2pt
              \vrule width#1pt height#2pt depth\dimen0}
\def\vR#1,#2;{\vrule height#1pt depth#2pt}
\def\blb#1#2#3#4#5
            {\hbox{\ifnum#2=0\hskip11.5pt
                   \else\ifnum#2=1\hr13,5;\hskip-1.5pt
                   \else\ifnum#2=2\hr13.5,6.5;\hskip-13.5pt
                                  \hr13.5,3.5;\hskip-2pt
                   \else\ifnum#2=3\hr13.7,8;\hskip-13.7pt
                                  \hr13,5;\hskip-13pt
                                  \hr13.7,2;\hskip-2.2pt\fi\fi\fi\fi
                   $#1$
                   \ifnum#4=0
                   \else\ifnum#4=1\hskip-9.2pt\vR22,-9;\hskip8.8pt
                   \else\ifnum#4=2\hskip-10.9pt\vR22,-8.75;\hskip3pt
                                  \vR22,-8.75;\hskip7.1pt
                   \else\ifnum#4=3\hskip-12.6pt\vR22,-8.5;\hskip3pt
                                  \vR22,-9;\hskip3pt
                                  \vR22,-8.5;\hskip5.4pt\fi\fi\fi\fi
                   \ifnum#5=0
                   \else\ifnum#5=1\hskip-9.2pt\vR1,12;\hskip8.8pt
                   \else\ifnum#5=2\hskip-10.9pt\vR1.25,12;\hskip3pt
                                  \vR1.25,12;\hskip7.1pt
                   \else\ifnum#5=3\hskip-12.6pt\vR1.5,12;\hskip3pt
                                  \vR1,12;\hskip3pt
                                  \vR1.5,12;\hskip5.4pt\fi\fi\fi\fi
                   \ifnum#3=0\hskip8pt
                   \else\ifnum#3=1\hskip-5pt\hr13,5;
                   \else\ifnum#3=2\hskip-5.5pt\hr13.5,6.5;
                                  \hskip-13.5pt\hr13.5,3.5;
                   \else\ifnum#3=3\hskip-5.7pt\hr13.7,8;
                                  \hskip-13pt\hr13,5;
                                  \hskip-13.7pt\hr13.7,2;\fi\fi\fi\fi

                   }}
\def\blob#1#2#3#4#5#6#7{\hbox
{$\displaystyle\mathop{\blb#1#2#3#4#5 }_{#6}\sp{#7}$}}
\def\up#1#2{\dimen1=33pt\multiply\dimen1by#1\hbox{\raise\dimen1\rlap{#2}}}
\def\uph#1#2{\dimen1=17.5pt\multiply\dimen1by#1\hbox{\raise\dimen1\rlap{#2}}}
\def\dn#1#2{\dimen1=33pt\multiply\dimen1by#1\hbox{\lower\dimen1\rlap{#2}}}
\def\dnh#1#2{\dimen1=17.5pt\multiply\dimen1by#1\hbox{\lower\dimen1\rlap{#2}}}
\def\rlbl#1{\kern-8pt\raise3pt\hbox{$\scriptstyle #1$}}
\def\llbl#1{\raise3pt\llap{\hbox{$\scriptstyle #1$\kern-8pt}}}
\def\elbl#1{\kern3pt\lower4.5pt\hbox{$\scriptstyle #1$}}
\def\lelbl#1{\rlap{\hbox{\kern-9pt\raise2.5pt\hbox{{$\scriptstyle #1$}}}}}

\def\whtd#1#2#3#4#5{\blob\circ#1#2#3#4{#5}{}}
\def\blkd#1#2#3#4#5{\blob\bullet#1#2#3#4{#5}{}}

\def\blkr#1#2#3#4#5{\blob\bullet#1#2#3#4{}{}\rlbl{#5}}

\def\ddeiiit#1.#2.#3.#4.#5.#6.#7.#8.{\hbox{$\vcenter{\hbox
       {\dynk \whtd0100{#1}\blkd1100{#3}%
       \up1{\blkr0001{#2}}\whtd1110{#4}\blkd1100{#5}\whtd1100{#6}%
       \blkd1100{#7}\whtd1000{#8}} }$}}
%
%

\def\dfil{\leaders\hbox to .1in{\hfil$\cdot\,$}\hfil\hskip -.005in}
\def\hs#1{\hbox to .5in{\dfil$#1$}}
\def\rs#1{\hbox to .5in{\hfil$\a_{#1}$}}
\def\label#1{\llap{$\scriptstyle #1$}}
\def\N{\hbox to .5in{\dfil}}
\def\Y{\hs{\B}}
\def\YY{\hs{\B\B}}
\newcount\n
\newcount\m
\def\y#1{\hs{\m=0\loop\ifnum\m<#1\advance\m by1{\mkern -1.5mu\B}\repeat}}
\def\ys#1.#2{\n=0\loop\ifnum\n<#2\advance\n by1\y#1\repeat}
\def\Ys#1{\n=0\loop\ifnum\n<#1\advance\n by1\Y\repeat}
\def\Ns#1{\n=0\loop\ifnum\n<#1\advance\n by1\N\repeat}
\def\As#1{\n=0\loop\ifnum\n<#1\advance\n by1\rs{\the\n}\repeat}
\def\row#1.#2{\hbox{\label{#1~}#2}}
%
%
\newdimen\bw
\newdimen\hbw
\newdimen\tw
\bw=8.2pt
\hbw=4.1pt
\def\ha{\hskip .4pt}
\tw=135pt
\def\bx{\vbox{\hrule\hbox to \bw{\vrule height 2.5pt%
\hfil\vrule}}}
\def\bxs#1.{\n=0\loop\ifnum\n<#1\advance\n by1\bx\repeat}
\def\nbx{\ha\hskip\bw\ha}
\def\p#1{\ha\vbox{\offinterlineskip\bxs#1.}\ha}
\def\q#1{\n=0\loop\ifnum\n<#1\advance\n by1\nbx\repeat}
\def\P{\p1}
\def\nt{\tline9.0.1.\hfil}
\def\bag#1{\vbox{\offinterlineskip
\hbox{#1}
\hrule
\hbox to \tw{\tick0.2.\nt\tick0.2.\nt\tick0.2.\nt\tick0.2.}
}}
\def\bagh#1{\bag{\hskip\hbw\ha#1\ha\hskip\hbw}}
%
%
%
\def\RF#1#2{\if*#1\ref#1{#2.}\else#1\fi}
\def\NRF#1#2{\if*#1\nref#1{#2.}\fi}
\def\refdef#1#2#3{\def#1{*}\def#2{#3}}
%
%
\def\ts{\hskip .16667em\relax}
\def\AJM{{\it Am.\ts J.\ts Math.\ts}}
\def\ADVM{{\it Adv.\ts Math.\ts}}

\def\IJMP{{\it Int.\ts J.\ts Mod.\ts Phys.\ts}}
\def\INVM{{\it Inv.\ts Math.\ts}}

\def\JP{{\it J.\ts Phys.\ts}}

\def\NP{{\it Nucl.\ts Phys.\ts}}
\def\PL{{\it Phys.\ts Lett.\ts}}

\def\PR{{\it Phys.\ts Rev.\ts}}

\def\TAMS{{\it Trans.\ts Amer.\ts Math.\ts Soc.\ts}}
\def\SJNP{{\it Sov.\ts J.\ts Nucl.\ts Phys.\ts}}
\def\Tahoe{Proceedings of the XVIII International Conference on
 Differential Geometric Methods in Theoretical Physics: Physics
 and Geometry, Lake Tahoe, USA 2-8 July 1989}
\def\Tahoe{Proc.~NATO
 Conference on Differential Geometric Methods in Theoretical
 Physics, Lake Tahoe, USA 2-8 July 1989 (Plenum 1990)}
\def\Zm{Zamolodchikov}
\def\AZm{A.\ts B.\ts \Zm}
\def\AlZm{Al.\ts B.\ts \Zm}
\def\me{P.\ts E.\ts Dorey}
\def\dur{H.\ts W.\ts Braden, E.\ts Corrigan, \me\ and R.\ts Sasaki}
%
%
\refdef\rAFZa\AFZa{A.\ts E.\ts Arinshtein, V.\ts A.\ts Fateev and
 \AZm, \PL {\bf B87} (1979) 389}
\refdef\rBa\Ba{H.\ts W.\ts Braden,
 \JP {\bf A25} (1992) L15}
\refdef\rBc\Bc{N.\ts Bourbaki, {\it Groupes et alg\`ebres de Lie} {\bf
 IV, V, VI,} (Hermann, Paris 1968)}
\refdef\rBCa\BCa{R. Brunskill and A. Clifton-Taylor, {\it English Brickwork}
 (Hyperion 1977)}
\refdef\rBCDSb\BCDSb{\dur,
`Aspects of perturbed conformal field theory, affine Toda field
theory and exact S-matrices',
\Tahoe}
\refdef\rBCDSc\BCDSc{\dur,
\NP {\bf B338} (1990) 689}
\refdef\rBCDSe\BCDSe{\dur,
 \NP {\bf B356} (1991) 469}
\refdef\rBCDSf\BCDSf{\dur, \NP {\bf B338} (1990) 689;
 \NP {\bf B356} (1991) 469}
\refdef\rBPZa\BPZa{A. A. Belavin, A. M. Polyakov and A. B. Zamolodchikov, \NP
 {\bf B241} (1984) 333}
\refdef\rCf\Cf{R.\ts Carter, {\it Simple Groups of Lie Type}, (Wiley
 1972)}
\refdef\rCn\Cn{R.\ts Carter, `Conjugacy classes in the Weyl groups',
{\it Comp.\ts Math.\ts} {\bf 25} (1972) 1}
\refdef\rCMe\CMe{P.\ts Christe and G.\ts Mussardo, \NP {\bf B330} (1990)
465}
\refdef\rDc\Dc{\me,
`Root systems and purely elastic S-matrices',
 \NP {\bf B358} (1991)~654}
\refdef\rDd\Dd{\me,
`Root systems and purely elastic S-matrices II',
 \NP {\bf B374} (1992)~741}
\refdef\rDg\Dg{\me,
`Partition Functions, Intertwiners and the Coxeter Element',
 preprint SPhT/92-053, hep-th/9205040; \IJMP {\bf A}, in press}
\refdef\rDGZa\DGZa{G.\ts W.\ts Delius, M.\ts T.\ts Grisaru and D.\ts Zanon,
`Exact S-matrices for non simply-laced affine Toda theories',
 \NP {\bf B282} (1992) 365}
\refdef\rDRa\DRa{\me\ and F.\ts Ravanini,
`Staircase models from affine Toda field theory',
 preprint SPhT/92-065, DFUB-92-09, hep-th/9206052, \IJMP {\bf A} in press}
\refdef\rDRb\DRb{\me\ and F.\ts Ravanini,
`Generalising the staircase models',
 preprint CERN-Th.6739/92, DFUB-92-21, NI\ts 92009, hep-th/9206052;
 \IJMP {\bf A} in press}
\refdef\rDZa\DZa{P.\ts Di\ts Francesco and J.-B.\ts Zuber, \NP {\bf B338}
 (1990) 602}
\refdef\rFb\Fb{M.\ts D.\ts Freeman,
\PL {\bf B261} (1991) 57}
\refdef\rFKMa\FKMa{P.\ts G.\ts O.\ts Freund, T.\ts Klassen and E.\ts Melzer,
 {\it Phys. Lett.} {\bf B229} (1989) 243}
\refdef\rFLOa\FLOa{A.\ts Fring, H.\ts C.\ts Liao and D.\ts I.\ts Olive,
   \PL {\bf B266} (1991) 82}
\refdef\rFOa\FOa{A.\ts Fring and D.\ts I.\ts Olive,
`The fusing rule and the scattering matrix of affine Toda theory',
 \NP {\bf B379} (1992) 429}
\refdef\rKa\Ka{M.\ts Karowski,
 `On the bound state problem in 1+1 dimensional field theories',
 \NP {\bf B153} (1979) 244}
\refdef\rKb\Kb{B.\ts Kostant, \AJM {\bf 81} (1959) 973}
\refdef\rKe\Ke{V.\ts Kac, Infinite dimensional Lie algebras, Third
Edition (Cambridge University Press 1990)}
\refdef\rKh\Kh{V.\ts G.\ts Kac,
`Infinite-Dimensional Algebras, Dedekind's $\eta$-Function,
Classical M\"obius Function and the Very Strange Formula',
\ADVM {\bf 30} (1978) 85}
\refdef\rKWa\KWa{V.\ts Kac and M.\ts Wakimoto, ``Exceptional Hierarchies
of Soliton Equations", {\it Proc. Symp. Pure Math.}\ts {\bf 49} (1989) 191}
\refdef\rOTa\OTa{D. I. Olive and N. Turok, {\it Nucl. Phys.} {\bf B215} (1983)
 470}
\refdef\rPc\Pc{V.\ts Pasquier, \NP {\bf B285} (1987) 162}
\refdef\rSf\Sf{F.\ts Smirnov,
 `Reductions of the sine-Gordon model as a perturbation of minimal
 models of conformal field theory',
 \NP {\bf B337} (1990) 156}
\refdef\rSg\Sg{N.\ts Sochen, \NP {\bf B360} (1991) 613}
\refdef\rSh\Sh{R.\ts Steinberg, \TAMS {\bf 91} (1959) 493}
\refdef\rSj\Sj{T.\ts A.\ts Springer,
 `Regular elements of finite reflection groups',
 \INVM {\bf 25} (1974) 159}
\refdef\rSBa\SBa{H.\ts Saleur and M.\ts Bauer, \NP {\bf B320} (1989)
591}
\refdef\rSWa\SWa{R. Shankar and E. Witten, \PR {\bf D17} (1978) 2134}
\refdef\rZa\Za{\AZm, `Integrable Field Theory from Conformal Field Theory',
 Proc. Taniguchi Symposium, Kyoto (1988)}
\refdef\rZb\Zb{\AZm, {\it Int. J. Mod. Phys.} {\bf A4} (1989) 4235}
\refdef\rZc\Zc{\AZm, {\it JETP Letters} {\bf 43} (1986) 730}
\refdef\rZd\Zd{\AZm, {\it Int. J. Mod. Phys.} {\bf A3} (1988) 743}
\refdef\rZe\Ze{\AZm, {\it Sov. Sci. Rev., Physics}, {\bf v.2} (1980)}
\refdef\rZf\Zf{\AZm, {\it Teor. Mat. Fiz.} {\bf 65} (1985) 347}
\refdef\rZp\Zp{\AZm, \SJNP {\bf 44} (1986) 529}
\refdef\rZq\Zq{\AZm,
``Renormalization group and perturbation theory about fixed points in
two-dimensional field theory",
\SJNP {\bf 46} (1987) 1090}
\refdef\rZz\Zz{\Za; {\it Int. J. Mod. Phys.} {\bf A4} (1989) 4235}
\refdef\rZZa\ZZa{\AZm\  and \AlZm, {\it Ann. Phys.}
 {\bf 120} (1979) 253}
\refdef\rWWa\WWa{G.\ts M.\ts T.\ts Watts and R.\ts A.\ts Weston,
 `$G_2^{(1)}$ affine Toda field theory. A numerical test of exact
 S-matrix results',
 \PL {\bf B289} (1992) 61}
\refdef\rKWb\KWb{H.\ts G.\ts Kausch and G.\ts M.\ts T.\ts Watts,
`Duality in quantum Toda theory and $W$-algebras',
\NP {\bf B386} (1992) 166}
\def\tilde{\widetilde}
\noblackbox
\Title{\vbox{\baselineskip12pt%
\hbox{NI\ts 92018}\hbox{hep-th/9212143}}}%
{Hidden geometrical structures in integrable
models$^{\scriptscriptstyle 1}$\footnote{}{\kern -15pt$^1$ Based on a
talk given at
the conference ``Integrable Quantum Field Theories" held in Como, Italy,
13-19 September 1992}}
\centerline{\titlerms Patrick Dorey}
\smallskip
{\baselineskip 12.5pt\footnotefont
\newt\centerline{\it and}
\centerline{CERN TH, 1211 Geneva 23, Switzerland}
\centerline{\tt dorey@surya11.cern.ch}
}
\vskip .6in
{\baselineskip 12.2pt\footnotefont\noindent
The bootstrap equations for the ADE series of purely elastic
scattering theories have turned out to be intimately connected with
the geometry of root systems and the Coxeter element. An informal
review of some of this material is given, mentioning also a couple of
other contexts -- the Pasquier models, and the simply-laced affine
Toda field theories -- where similar structures are encountered. The
relevance of twisted Coxeter elements is indicated, and a construction
of these elements inspired by the twisted foldings of the affine Toda
models is described.
}

\Date{November 1992}

\nwsec{An example}
To provide some motivation, this first section is devoted to the scaling
region of the 2D Ising model in a magnetic field. However, the
physical relevance of the discussion will not be immediately apparent.

So, the starting point is a certain set $\Phi$ of $240$ vectors
(`roots') in eight-dimensional space, each having length-squared 2,
known as the $E_8$ root system. For each $\a\in\Phi$, there is a
reflection $r_{\a}$ in the 7-dimensional hyperplane
orthogonal to $\a$, and a key feature of $\Phi$ is that
these reflections generate a {\it finite} group, $W$ say. In symbols,
\eqn\weyldef{\eqalign{&~~r_{\a}(x)=x-2{\a.x\over \a^2}\a\,;\cr
  &|\VEV{\{r_{\a}\}_{\a\in\Phi}}|=|W|<\infty\,.\cr}}
This group is none other than the Weyl group of the Lie algebra
$E_8$, more abstractly defined inside the group of inner automorphisms
of the algebra as the
quotient of the normaliser of a Cartan subalgebra (those inner
automorphisms that map the subalgebra into itself) by the centraliser
(those that leave it pointwise fixed). But for now this fact is best
forgotten, as the geometrical characterisation just given is going to
be the relevant one. Two further properties of $W$ will be
needed: first, $W$ leaves the set $\Phi$ invariant, and
second, $W$ can be generated by the reflections for a subset
$\Delta\subset\Phi$ of just eight so-called simple roots. Actually, there are
$240$ such subsets, but they are all
equivalent in the sense of being conjugate under $W$, so we can focus
on one, and label its elements $\a_1,\a_2,\dots ,\a_8$, and the
corresponding reflections $r_1,r_2,\dots ,r_8$.
All of $W$ is encoded in the mutual inner products of these eight
vectors, given equivalently by the Cartan matrix:
\eqn\cmatr{C_{ab}=2{\a_a.\a_b\over\a_b^2}}
or the Dynkin diagram:
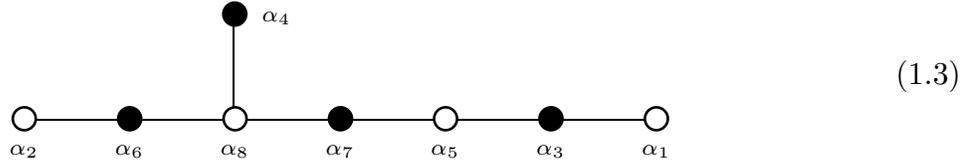
\eqn\ddiag{\ddeiiit \a_2.\a_4.\a_6.\a_8.\a_7.\a_5.\a_3.\a_1.}
If two simple roots are joined by a single line, then they have inner
product $-1$; otherwise, it is $0$. The only unusual feature of
\ddiag\ is the two-colouring: to maintain an element of
suspense, this will not be explained
immediately, but rather used to define a particular
element $w$ of the Weyl group, as follows:
\eqn\coxdef{w=r_3r_4r_6r_7r_1r_2r_5r_8\,.}
Those who have encountered these things before might recognize $w$ as
a Coxeter element, in a Steinberg ordering -- that is, a product of
all the generating reflections, in an ordering such that the
`white' reflections act first, followed by the `black' ones.
As an -- at the moment completely
unmotivated -- exercise, consider computing the repeated action of
$w^{-1}$ on, say, the simple root $\a_1$.
Each $r_a$ squares to the identity, so
$w^{-1}=r_8r_5r_2r_1r_7r_6r_4r_3\,$. By \weyldef\ and
\cmatr, the action of a simple reflection $r_a$ on any simple
root $\a_b$ is just $r_a\a_b=\a_b-C_{ba}\a_a\,$.
In terms of the Dynkin diagram, $r_a$ negates $\a_a$, adds $\a_a$ to each
$\a_b$ joined to $\a_a$ by a link, and leaves all of the other
simple roots unchanged.
So, working in from the right of $w^{-1}\,$,
$r_3$ sends $\a_1$ to $\a_1{+}\a_3$, and then
$r_4$, $r_6$ and $r_7$ all leave these two roots
alone. Then comes $r_1$, which negates $\a_1$ while at the same time
transforming $\a_3$ into $\a_3{+}\a_1$, so that the total is now just
$\a_3$. Next is $r_2$, which does nothing, and then $r_5$ on $\a_3$
produces $\a_3{+}\a_5$. These last two roots are both orthogonal to $\a_8$,
and hence are unchanged by the final action of $r_8$. Thus
\eqn\wact{w^{-1}\a_1=\a_3+\a_5\,.}
Nothing apart from growing tedium prevents us from carrying on further,
and calculating $w^{-2}\a_1$, $w^{-3}\a_1$ and so on. The sequence must
ultimately repeat, since $|\Phi|$ is finite; in fact, $\a_1$ appears
on its own again after $30$ steps. The first $14$ of these steps
are shown in the following table, where the multiplicity of
$\a_a$ in $w^{-p}\a_1$ is given by the number of
blobs ($\B$) that appear in the $a\th$ position of the $p\th$ row.
\medskip
\eqn\tableone{\vcenter{\hbox{Images of $\a_1\!$ under $w^{-1}\,$:}
\smallskip
\baselineskip=9pt
\row14.{\Y\N\Y\Ns5}
\row13.{\Ns4\Y\N\Y\N}
\row12.{\Ns3\Y\N\Ys3}
\row11.{\N\Y\Ns3\Ys3}
\row10.{\Ns2\Ys3\N\Ys2}
\row9.{\Y\N\Y\N\Ys4}
\row8.{\N\Y\N\Ys5}
\row7.{\Ns3\Y\N\Ys2\YY}
\row6.{\N\Ys2\N\Ys4}
\row5.{\Y\N\Ys3\N\Ys2}
\row4.{\Ns4\Ys4}
\row3.{\N\Y\N\Y\N\Ys3}
\row2.{\Ns6\Ys2}
\row1.{\Ns2\Y\N\Y\Ns3}
\row0.{\Y\Ns7}
\hbox{\As8}
}}
\medskip\noindent
This was a deceptively easy case -- the orbits
become much more complicated, although they always contain exactly $30$
elements. Worst of all is that for $\a_8$:
\medskip
\eqn\tabletwo{\vcenter{\hbox{Images of $\a_8\,$:}
\smallskip
\baselineskip=9pt
\row14.{\Ns3\Y\N\Ys3}
\row13.{\N\Ys4\ys2.3}
\row12.{\Ys2\ys2.4\ys3.2}
\row11.{\Ys2\ys2.2\ys3.2\ys4.2}
\row10.{\Y\ys2.2\ys3.2\ys4.2\ys5.1}
\row9.{\Y\ys2.2\ys3.2\ys4.1\ys5.1\ys6.1}
\row8.{\Y\ys2.1\ys3.2\ys4.2\ys5.1\ys6.1}
\row7.{\ys2.2\ys3.2\ys4.2\ys5.1\ys6.1}
\row6.{\Y\ys2.2\ys3.1\ys4.2\ys5.1\ys6.1}
\row5.{\Y\ys2.2\ys3.2\ys4.2\ys6.1}
\row4.{\Y\ys2.3\ys3.2\ys4.1\ys5.1}
\row3.{\Ys2\ys2.2\y3\y2\y3\y4}
\row2.{\Ys4\ys2.3\ys3.1}
\row1.{\N\Y\N\Ys4\y2}
\row0.{\Ns7\Y}
\hbox{\As8}
}}
\medskip\noindent
To give an example, $w^{-7}\a_8=
2\a_1{+}2\a_2{+}3\a_3{+}3\a_4{+}4\a_5{+}4\a_6{+}5\a_7{+}6\a_8\,$.
This root is made from a total of
$2{+}2{+}3{+}3{+}4{+}4{+}5{+}6=29$ simple roots, which is equivalent
to saying that its `height' is $29$. In fact this is the largest
possible height in $E_8$, and furthermore it occurs just once for a
given choice of simple roots.
Hence $w^{-7}\a_8$ should be equal to $\psi$,
the highest root of $E_8$, and entering its coefficients on the Dynkin
diagram \ddiag\ to see that this is indeed the case provides a quick
check on the calculation. One last point before moving on: in
compiling these two tables, there was no need to invent a notation for the
negative of a simple root -- an `antiblob', perhaps -- as all the roots
from $0$ to $14$ were {\it positive}-linear combinations of the simple
roots. This will be relevant later.

But first, Zamoldchikov's ideas about the scaling region of
the Ising model in a magnetic field. At criticality, the continuum
limit of the model is well-known to be described by a $c{=}\half$
conformal field theory\ts\RF\rBPZa\BPZa; Zamolodchikov's
proposal\ts\RF\rZz\Zz\ was to probe the neighbourhood of this point
via a study of actions
\eqn\pertact{S_{pert}=S_{CFT}+\lambda\!\!\int\!\!{\rm d^2}x\,\phi(x)\,,}
where $S_{CFT}$ is a notional action for the conformal theory, inside
of which $\phi$ is some (relevant, spinless) field, the coupling
constant $\lambda$ ensuring that all the dimensions match up. There
are just two possibilities for this simplest form of $S_{pert}$ in the
case of the Ising model, one for each of the two spinless relevant fields
in the $c{=}\half$ operator algebra. These are usually
labelled $\sigma$ and $\epsilon$, and have dimensions $({1\over
16},{1\over 16})$ and $(\half,\half)$ respectively. From their
identifications with the scaling limits of the
local magnetisations (spins) and energy densities on the lattice,
perturbing by $\sigma$ corresponds to switching on a magnetic
field, and perturbing by $\epsilon$ to changing the temperature $T$
away from its critical value $T_c$.

Zamolodchikov was able to extract
much non-perturbative information about the theories defined
through \pertact, essentially because the perturbative
expansions for certain quantities of interest truncate after finitely
many terms. In particular, his `counting argument' led to
the conclusion that both the $\epsilon$ and $\sigma$ perturbations
of the Ising model preserve certain higher-spin conserved
charges, and hence are integrable.
For the $\sigma$ perturbation, these spins are the numbers
$1,7,11,13,17,19,23,29$, repeated modulo $30$.

The next stage is to study the long-distance behaviour of the theory
described by $S_{pert}$.  In the ultraviolet, the theory is
well-approximated by the original conformal theory; in the infrared, the
model might be massive, or alternatively it might undergo a
crossover to another conformal field theory. However, in the latter
case the central charge of the infrared model is constrained by
Zamolodchikov's $c$-theorem\ts\RF\rZc\Zc\ to be less than that of the
original conformal theory. Here, $c$ started at $\half$, below
which there is no unitary central charge, so this possibility is ruled
out. Hence the infrared limit is massive, something that should in any
case have been expected from knowledge of the phase diagram of the
lattice model. Now after a Wick rotation, this massive model
can also be examined in
Minkowski space, where it should have an S-matrix. Better than that,
since the perturbation was integrable, the S-matrix should be
factorizable\ts\RF\rZZa\ZZa, and perhaps even findable\dots

Easiest of all would be for the S-matrix to be diagonal -- the theory
would then consist of a collection of scalar particles, perhaps with
different masses, which never mix under scattering. In this way, the
joys/sorrows of the Yang-Baxter equation would be avoided, and the
$2{\rightarrow}2$ S-matrix would boil down to a collection of meromorphic
functions, one for each pair of particle types in the model. The
critical Ising model being the simplest unitary conformal field theory,
one might expect its integrable deformations
to have S-matrices of this simplest possible form. Indeed, at
$T{\neq}T_c$ (the $\epsilon$ perturbation) the model is a
theory of free massive fermions, which certainly have a diagonal
S-matrix. Perturbing by $\sigma$, things are not quite so simple,
but in fact the S-matrix is again diagonal.
A sequence of ingenious arguments, combining general principles with
certain physical inputs specific to the Ising model, led
Zamolodchikov to propose the following expression for the S-matrix
element for the scattering of two of the lightest particles, as a
function of their relative rapidity $\t\,$:
\eqn\sii{
S_{11}=-\usbl{2}\usbl{10}\usbl{12}\usbl{18}\usbl{20}\usbl{28}\,,}
where an abbreviated notation has been used:
\eqn\usbldef{
\usbl{x}\equiv{\sinh\bigl({\theta\over 2}+{i\pi x\over 60}\bigr)\over
        \sinh\bigl({\theta\over 2}-{i\pi x\over 60}\bigr)}~.}
Note, $\usbl{x}$ has a pole at $i\pi x/30$, so \sii\
exhibits the poles in $S_{11}(\t)$ between $0$ and $i\pi$.

How to proceed from here? The key notion is that of a {\it bootstrap
equation}\ts\NRF\rKa\Ka\refs{\rKa,\rZz}. Assume that an S-matrix
element $S_{ab}(\t)$ has a simple pole at $\t{=}iU^c_{ab}$, say,
with residue a positive-real multiple of $i$. Transforming back to the
Mandlestam variable $s=(p_a{+}p_b)^2$ reveals a positive-residue pole
below threshold, which should still correspond to the formation of a
(forward channel)
bound state. If this is assigned the charge-conjugated label $\bar c$,
then its presence as a bound state in $a~b$ scattering corresponds to
the non-vanishing of the three-point coupling $C^{abc}$. Kinematic
considerations (the conservation of momentum and the fact that $a$,
$b$ and $c$ should all be on-shell at the pole) then give
a relation
between the particle masses and the `fusing angle' $U^a_{bc}\,$:
\eqn\angdef{m_c^2=m_a^2+m_b^2+2m_am_b\cos U^c_{ab}\,.}
That the process happens below threshold requires $m_c<m_a{+}m_b$,
consistent with $U_{ab}^c$ being real. Hence all relative
momenta are euclidean, and the process can be drawn in the plane:
\eqn\scattpic{\eqalign{
&\kern .1pt\llap{$U^b_{ac}~~$}\clap{$\titlefont\Big\uparrow$}%
\raise3.1ex\clap{$\scriptstyle \bar c$}\rlap{$~~\,U^a_{bc}$}\cr
\noalign{\vskip -2pt}
\llap{$\titlefont\nearrow$}&\lower1ex\llap{$\scriptstyle a~~~$}%
\lower2ex\clap{$U_{ab}^c$}\lower1.1ex\rlap{$\scriptstyle ~~~b$}%
\rlap{$\titlefont\nwarrow$}%
\cr}}
Rotating the diagram by $\pm 2\pi/3$ gives pictures for $a~c$ and
$b~c$ scattering, and the corresponding fusing angles have also been
marked in; the triplet of angles satisfies
\eqn\Urel{U^a_{bc}+U^b_{ac}+U^c_{ab}=2\pi\,.}
Now imagine that \scattpic\ is just part of a larger diagram,
involving at least one further particle $d$.
Depending on its impact parameter, the world-line of particle $d$
will either cross those of $a$ and $b$, or else that of $c$,
corresponding to the interaction with $d$ happening either before or
after the fusing of $a$ with $b$. (The idea is that the momenta of all
particles have been slightly `smudged' about the values given, so that
a description in terms of localised wavepackets propagating freely in
between interactions is valid.) Now our system is integrable, which
among other things means that the values of impact parameters should
be irrelevant -- and so, irrespective of what happens elsewhere in the
larger diagram, we should certainly equate the two possible
contributions to the part of the amplitude from particle $d$.
This translates into a bootstrap relation between S-matrix elements,
expected to hold whenever a positive-residue simple pole in
$S_{ab}(\t)$ has led us  to deduce the non-vanishing of the
three-point coupling $C^{abc}\,$:
\eqn\bootsm{S_{d\bar c}(\t)
 =S_{da}(\t{-}\bar U^b_{ac})S_{db}(\t{+}\bar U^a_{bc})\,,}
where $\bar U{=}\pi{-}U\,$.
(To see why, shift the diagram \scattpic\ to
the frame for which the rapidity of particle $c$ is zero, whereupon
the rapidities of $a$ and $b$ become $-\bar U^b_{ac}$ and
$\bar U^a_{bc}$ respectively.  Then imagine $d$ to traverse the
picture either above or below the fusing, and equate the results.)
This `derivation'
is in the spirit of a discussion of the Yang-Baxter equation given
by Shankar and Witten\ts\RF\rSWa\SWa, compared to which there are
(at least) two further dodgy points: first, some rapidities are
necessarily unphysical, and second, the way in which particles
$a$ and $b$ fuse to form $c$ has been left imprecise. The second
objection is potentially the more serious, especially since
there do appear to be situations -- for example, breather-breather
scattering below the two-breather threshold in the sine-Gordon
model\ts\RF\rSf\Sf, or certain amplitudes in
the non simply-laced affine Toda
theories\ts\RF\rDGZa\DGZa\ -- where the competing presence of a number
of anomalous threshold poles at the same point may prevent the
direct identification of a simple pole in the S-matrix with a
single bound state. The bootstrap equation \bootsm\ is therefore best
taken as a `working axiom', to be re-examined in the event
that it contradicts other information. For the case in hand, there don't
seem to be any problems.

The S-matrix element \sii\ has forward-channel poles from the
blocks $\usbl{2}$, $\usbl{12}$ and $\usbl{20}$, with fusing angles
of $\pi/15$, $2\pi/5$ and $2\pi/3$ respectively. Via \angdef, these
correspond to bound state masses $m_3{=}m_1\sin 2\pi/30$,
$m_2{=}2m_1\sin 3\pi/10$ and $m_1$, where $m_1$ is the common mass of
the two incident particles. The simplest possibility for the third of
these is just another copy of the (then
self-conjugate) particle $1$, implying $C^{111}{\neq}0$. Using this
`$\phi^3$' property in the bootstrap equation then implies a rather
strong constraint on $S_{11}(\t)\,$:
\eqn\siiboot{S_{11}(\t)=S_{11}(\t{-}i\pi/3)S_{11}(\t{+}i\pi/3)\,.}
It is instructive to check the intricate cancellations which
ensure that the expression \sii\ passes this test. But to get something
new, the other bound state poles should be exploited,
which have been provisionally associated with the second and
third lightest particles in the model. For the $1~1\rightarrow2$ fusing,
the bootstrap equation predicts
\eqn\sitboot{\eqalign{
S_{12}(\t)&=S_{11}(\t{-}i\pi/5)S_{11}(\t{+}i\pi/5)\cr
          &=\usbl{6}\usbl{8}\usbl{12}\usbl{14}\usbl{16}\usbl{18}%
            \usbl{22}\usbl{24}\,,\cr}}
while $1~1\rightarrow 3$ yields
\eqn\sitiboot{\eqalign{
S_{13}(\t)&=S_{11}(\t{-}i\pi/30)S_{11}(\t{+}i\pi/30)\cr
          &=\usbl{1}\usbl{3}\usbl{9}\usbl{11}^2\usbl{13}\usbl{17}%
            \usbl{19}^2\usbl{21}\usbl{27}\usbl{29}\,.\cr}}
The forward channel poles are proliferating: between them, $S_{12}$
and $S_{13}$ have eight, five of which are explicable in terms of
bound states of types $1$, $2$ and $3$, while the remaining three
require the introduction of two new species, call them $4$ and $5$.
The poles also imply the non-vanishing of further three-point
couplings: $C^{122}$, $C^{123}$ and $C^{124}$ from $S_{12}$, and
$C^{134}$ and $C^{135}$ from $S_{13}$. This information can be fed
back into \bootsm\ to give more bootstrap equations, which on the one
hand provide further tests for the existing set of S-matrix elements,
and on the other add to this set by predicting new ones.

Now iterate! The procedure just outlined is sufficiently well-defined
that there is nothing (apart, again, from growing tedium) to prevent
its continuation, resulting in an increasing collection of particles,
S-matrix elements and non-vanishing three-point couplings, all bound
together by the bootstrap equations \bootsm. But it is worth
persevering: as Zamolodchikov\ts\rZz\ discovered, quite remarkably the
process closes in on itself after a total of eight particle types have
been encountered. Tables of the $36$ meromorphic functions
which form their mutual two-particle S-matrix elements (recall,
$S_{ab}{=}S_{ba}$), and the corresponding non-vanishing three-point
couplings, were given in
refs.\ts\NRF\rBCDSb\BCDSb\NRF\rBCDSc\BCDSc\refs{\rBCDSb,\rBCDSc},
and it is a finite, though lengthy, task to verify that the system is
closed: all forward-channel poles are satisfactorily
explained, and all the corresponding bootstrap equations are
obeyed.

However, all this turns out to be
unnecessary\ts\RF\rDc\Dc. The iteration of the bootstrap equations
for the spin-perturbed Ising model is simply a more complicated way
of doing the Weyl group computation presented in the first half of
this section, while the fact that these equations are implied by the
pole structure of the very functions that they constrain,
when viewed from this perspective, follows
from simple properties of the $E_8$ root system. Some
sort of connection was perhaps to have been expected, given that (a)~the
$c{=}\half$ conformal field theory can be realised as the coset model
$E_8^{(1)}{\times}E_8^{(1)}/E_8^{(2)}$, within which the field
$\sigma$ corresponds to the branching $(id,id,adj)\,$; (b)~the spins
of the conserved charges given earlier are precisely the exponents
of $E_8$, repeated modulo the Coxeter number; and (c)~the masses
$m_1, m_2,\dots, m_8$ can be formed into the Perron-Frobenius
eigenvector for the incidence matrix of the $E_8$ Dynkin
diagram\ts\NRF\rFKMa\FKMa\refs{\rFKMa,\rBCDSb}. Nevertheless, the way
it works is quite striking, and this introduction
finishes with a visual demonstration of the result.

The key is to rewrite the S-matrix elements in a slightly different
way. Notice first that in all three S-matrix elements given above,
each block $\usbl{x}$ with $x$ not equal to $2$ or $28$ can be paired
off with another block, either $\usbl{x{-}2}$ or $\usbl{x{+}2}$.
Even the blocks $\usbl{2}$ and $\usbl{28}$ in $S_{11}(\t)$ can
be made to obey this rule, if they are paired formally with the `dummy'
blocks $\usbl{0}{\equiv}1$ and $\usbl{30}{\equiv}-1$. Since this
feature persists for all of the other S-matrix elements, it is
possible to economise on the previous formulae by introducing a new,
larger, building block\ts\rBCDSc:
\eqn\ubldef{\ubl{x}=\usbl{x{-}1}\usbl{x{+}1}\,.}
The earlier expressions become:
\eqn\newform{\matrix{
S_{11}&=&\ubl{1}\ubl{11}\ubl{19}\ubl{29}\hfill&
 =&\bag{\P\q4\P\q3\P\q4\P}\cr
S_{12}&=&\ubl{7}\ubl{13}\ubl{17}\ubl{23}\hfill&
 =&\bag{\q3\P\q2\P\q1\P\q2\P\q3}\cr
S_{13}&=&\ubl{2}\ubl{10}\ubl{12}\ubl{18}\ubl{20}\ubl{28}&
 =&\bagh{\P\q3\P\P\q2\P\P\q3\P}\cr}}
The `brick wall' notation for S-matrix elements, used here and
in ref.\thinspace\RF\rDd\Dd, represents each factor
$\ubl{x}$ in a product by a brick
$\vbox{\hbox to 12pt{\hfil\P\hfil}\hrule}\,$, centred
at $x$. The connection with the Weyl group data compiled earlier can
be seen by taking these three pieces of wall, rotating them by
$90^{\circ}$, and comparing them with the first three columns of
\tableone. To be sure that this isn't a coincidence, the
scattering amplitudes of the heaviest particle, $8$, can also
be checked:
\def\crv{\cr\noalign{\vskip 3pt}}
\eqn\moreform{\matrix{
S_{81}&=&\bag{\q2\P\P\P\P\P\p2\P\P\P\P\P\q2}\crv
S_{82}&=&\bag{\q1\P\P\P\p2\p2\p2\p2\p2\p2\p2\P\P\P\q1}\crv
S_{83}&=&\bagh{\q1\P\p2\p2\p2\p2\p3\p3\p2\p2\p2\p2\P\q1}\crv
S_{84}&=&\bagh{\P\P\p2\p2\p3\p3\p3\p3\p3\p2\p2\P\P}\crv
S_{85}&=&\bag{\q1\P\p2\p3\p3\p3\p4\p4\p4\p3\p3\p3\p2\P\q1}\crv
S_{86}&=&\bagh{\P\p2\p2\p3\p4\p4\p4\p4\p4\p4\p3\p2\p2\P}\crv
S_{87}&=&\bagh{\P\p2\p3\p4\p4\p5\p5\p5\p5\p4\p4\p3\p2\P}\crv
S_{88}&=&\bag{\P\p2\p3\p4\p5\p6\p6\p6\p6\p6\p5\p4\p3\p2\P}\cr}}
The functions represented here have become quite complicated -- for
example,
\eqn\see{S_{88}=\ubl{1}\ubl{3}^2\ubl{5}^3\ubl{7}^4\ubl{9}^5%
\ubl{11}^6\ubl{13}^6\ubl{15}^6\ubl{17}^6\ubl{19}^6\ubl{21}^5%
\ubl{23}^4\ubl{25}^3\ubl{27}^2\ubl{29}}
Nevertheless, every feature is perfectly reproduced by the Coxeter
orbit of $\a_8$, as listed in table~\tabletwo.
This is part of the `hidden geometrical structure'
advertised in the title, and the next section will outline some of the
uses to which it can be put.

\nwsec{More details}
The picture outlined above is not special to $E_8\,$: in fact, it
is found for any simply-laced Lie algebra $g=A$, $D$ or $E$.
To explain the general formalism, some more notation is needed:
this is the task of the first part of this section. For more on the
mathematical background, see
refs.\ts\NRF\rSh\Sh\NRF\rKb\Kb\NRF\rBc\Bc\NRF\rCf\Cf\refs{\rSh{--}\rCf}.

Most of the definitions given above can be carried straight over, on
replacing $8$ by $r$, the rank of $g$, and $30$ by $h$, the Coxeter
number. However,
although the two Coxeter orbits exhibited above were both based on
simple roots, this is not always possible, nor is it the most
convenient choice. It seems best to follow the treatment of
Kostant\ts\rKb: if the chosen Coxeter element is
\eqn\coxgen{w=r_1r_2\dots r_r\,,}
then, for $a=1\dots r$, define
\eqn\repdef{\phi_a=r_rr_{r-1}\dots r_{a+1}\a_a\,,}
the simple reflections acting in reverse order on the simple root
$\a_a$. A useful property of these roots is their
relation to the fundamental weights:
\eqn\wtrel{\phi_a=(1-w^{-1})\lambda_a\,.}
(This can be checked using the duality between the weights
$\lambda_a$ and the simple co-roots $\a^{\vee}_a\equiv (2/\a_a^2)\a_a$,
which implies, via \weyldef, that $r_a\lambda_b=
\lambda_b{-}\delta_{ab}\a_b\,$.) The eigenvalues of $w$ are $\exp(2\pi
is/h)$, where $s$ runs over the exponents of $g$; therefore, $1$
is not an eigenvalue, and so $R\equiv(1{-}w^{-1})^{-1}$ is
well-defined. Now the highest weights of different representations
lie in different Weyl-orbits, and in particular the fundamental weights
lie in distinct orbits of the Coxeter element
$w$. Translated via the action of $R$ into a statement about roots,
this implies that the $\phi_a$ all lie in different orbits of $W$.
Hence, defining $\Gamma_a=\{w^p\phi_a\}_{p=0}^{h-1}$
to be the orbit of $\phi_a$ under $w$, all of these orbits are distinct.
Inside $\Phi$, all orbits of a Coxeter element have exactly $h$
elements (note, the same statement is
not always true for weights, though via \wtrel\ it does hold for the
$\lambda_a$). Thus the union of the $\Gamma_a$ contains $r.h$ roots,
which is in fact the full set (recall for $E_8$ we had
$|\Phi|{=}240{=}8.30\,$), and all of the orbits of $w$ have
been captured in this way. Furthermore, the $\phi_a$ have a
distinguished position in the orbits: they are the (unique) positive
roots which become negative under the action of $w$\thinspace\rKb. (Recall
that the positive roots $\Phi^+$ are the positive-linear combinations
of the simple roots, the remainder of $\Phi$ being $\Phi^-$, the
negative-linear combinations: $\Phi=\Phi^+\cup\Phi^-$.)

It will also be convenient to assume that $w$ is written in the
`Steinberg ordering'\thinspace\rSh\ mentioned earlier. If the Dynkin
diagram of $g$ has been two-coloured as in \ddiag, then this requires
that
\eqn\wdef{w=\prod_{\B'}r_{\B'}\prod_{\W'}r_{\W'}=\wB\wW\,,}
where the symbol $\B'$ stands for an arbitrary black, and
$\W'$ an arbitrary white, simple root. (We could swap black and white;
this would just send $w$ to its inverse.) Note that since roots of
like colour are orthogonal, their reflections commute and as a
consequence $\wB^2=\wW^2=1\,$, and $w$ has been written as a product of
two involutions.  This looks to be a rather
special choice, but in fact it isn't: Carter\ts\RF\rCn\Cn\ has shown
that {\it any} element of the Weyl group can be written as the product
of two involutions as in \wdef, and that the Coxeter elements are exactly
those for which the inner products of the roots defining the two
involutions have mutual inner products given by the Dynkin diagram of
the algebra. Thus rather than thinking that a particular choice of Coxeter
element has been made, it could equally be said that the set of simple
roots has been changed at this point, to one better-suited to current
needs. In any event, with \wdef\ in place, an integer $u(\a,\b)$ can
be defined modulo $2h$ for each pair of roots $\a,\b\in\Phi\,$:
\eqn\udef{\eqalign{u(\a,\b)=-u(\b,\a)&\qquad u(w\a,\b)=u(\a,\b)+2\cr
u(\phi_{\B},\phi_{\B'})=u(\phi_{\W},\phi_{\W'})=0&\qquad
u(\phi_{\W},\phi_{\B})=1.\cr}}
As explained in \rDd, this definition has the geometrical meaning that
$\pi u(\a,\b)/h$ is the signed angle between the projections of $\a$ and
$\b$ into the $\exp(2\pi i/h)$ eigenspace of $w$ -- in terms of which,
it has a sense independent of the Steinberg ordering \wdef.

At last,
enough notation is in place to give the formula
which encodes the observations of the last section:
\eqn\Sabdef{S_{ab}=\prod_{\b\in\G^+_b}\{1+u(\phi_a,\b)\}^{(\l_a,\b)},}
where $\G^+_b=\G_b\cap\Phi^+$ is the positive part of the $w$-orbit
$\G_b$. To see that that this really does come to the same thing as
before, note
first that for $w$ given by \wdef, the definition \repdef\ reduces to
\eqn\throot{\phi_{\B}=\wW\alpha_{\B}\,,\qquad \phi_{\W}=\a_{\W}\,,}
and that both $\a_1$ and $\a_8$ were coloured white on \ddiag.
Also, the weights $\l_a$ are dual to the simple roots $\a_b$ in the
simply-laced cases, so that the exponent in \Sabdef\ is simply
counting the number of times that $\a_a$ appears in the expansion of
$\b$ in simple roots. That
all the roots shown in \tableone\ and \tabletwo\ were positive-linear
combinations of simple roots corresponds to the root $\b$ running over
only positive roots in \Sabdef. Finally, the colour-dependent
difference of $\pm 1$ in the value of
$u(\phi_a,\phi_b)$ implied by the last line of \udef\ exactly
accounts for the small shifts in some of the brick walls in \newform\
and \moreform\ as compared to the orbit tables \tableone\ and
\tabletwo\ -- for example, the fact that $S_{13}$ in \newform\ is
symmetrically placed relative to $S_{12}$, even though the second and
third columns of \tableone\ did not line up quite so nicely.
Physically, this is necessary to ensure that \Sabdef\ has the correct
crossing symmetry, and that the poles turn up in the correct
places.

Assuming that \Sabdef\ is the correct general formula, the first task
should be to examine its analytic structure.
The following basically follows~\rDd; an alternative discussion
can be found in~\RF\rFOa\FOa.
As $\b$ runs through $\G_b$, $1{+}u(\phi_a,\b)$ remains between $0$ and
$h$, a fact which can be seen geometrically\thinspace\RF\rDg\Dg, or
alternatively traced back to $\phi_b$ being the only positive
root in $\G_b$ to become negative on the action of $w$, together
with certain symmetries of the orbit between positive and negative
roots\ts\NRF\rFOa\FOa\refs{\rDd,\rFOa}.
{}From the definitions \usbldef, \ubldef\ of the blocks $\ubl{x}$, $S_{ab}$
might therefore have poles at any of the points $-i\pi/h\leq
i\pi u(\phi_a,\b)/h\leq i\pi{+}i\pi/h$, but it is possible to show (see,
for example, sect.~3 of \rDg) that any blocks which could contribute
poles at the extremal locations, outside the physical strip $0\leq
{\rm Im}\t\leq\pi$, are necessarily raised to the power zero in
\Sabdef. Hence in looking for poles in $S_{ab}$, attention can be
restricted to points $i\pi u(\phi_a,\b)/h$ inside the physical strip.
Remembering from \udef\
that $u(\phi_a,w\b)=u(\phi_a,\b){-}2$, it follows from \ubldef\ that
a such a pole can only receive direct
contributions from the blocks associated with $\b$ and
$w\b$, and that all other blocks only multiply this by a
positive real amount, which can be ignored for bootstrap purposes.
That is, near $i\pi u(\phi_a,\b)/h\equiv\t_0$ the relevant part of
\Sabdef\ is
\eqn\Sform{\eqalign{S_{ab}&=\dots\ubl{u(\phi_a,\b)-1}^{(\l_a,w\b)}
 \ubl{u{\phi_a,\b}+1}^{(\l_a,\b)}\dots\cr
&\approx \dots \({i\over\t-\t_0}\)^{(\l_a,w\b)}
\({-i\over\t-\t_0}\)^{(\l_a,\b)}\dots\cr}}
where in the second line the dominant contributions
at $\t_0$ have been extracted using \ubldef. Therefore the
residue is a positive-real multiple of
\eqn\respole{
 i^{(\l_a,w\b)-(\l_a,\b)}=i^{-((1-w^{-1})\l_a,\b)}
 =i^{-(\phi_a,\b)},}\nobreak
where \wtrel\ was used in the second equality.
In terms of the pictures \newform\ and \moreform, a $+i$
residue for the leading singularity occurs whenever there is a
downhill step in the wall of bricks, reading from left to right.
Empirically, this rule is also correct for assigning forward/crossed
channels to the odd-order poles of higher orders, and the
treatment in terms of root systems does not notice the distinction.
The way in which this works in perturbation theory is however
quite complicated\thinspace\RF\rBCDSe\BCDSe. But ignoring these
subtleties, and continuing to focus on the forward-channel poles (the
others can be treated in a similar way),
there is only one way that a residue of $+i$ can emerge from \respole:
we must have
\eqn\fwrd{(\phi_a,\b)=-1\,.}
In turn, this holds if and only if $\phi_a{+}\b=-\g\,$, say, is
another root. For
the `if', note that $\phi_a{+}\b=-\g\in\Phi$ implies
$2=\g^2=\phi_a^2{+}2(\phi_a,\b){+}\b^2$, and hence $(\phi_a,\b)=-1$,
all simply-laced roots having length-squared $2$. Conversely,
if \fwrd\ holds then, from \weyldef, $\phi_a{+}\b=r_{\b}\phi_a\,$,
and is therefore a root by the closure of $\Phi$ under $W$.
The situation in $\R^r$ can be drawn as follows:
\eqn\rootpic{\eqalign{
&\kern .1pt\clap{$\titlefont\Big\uparrow$}%
\raise3.4ex\clap{$\scriptstyle ~~-\g=\phi_a+\b$}\cr
\noalign{\vskip -2pt}
\llap{$\titlefont\nearrow$}&\lower1.4ex\llap{$\scriptstyle\phi_a~\,$}%
\lower1.4ex\rlap{$\scriptstyle ~~\b$}%
\rlap{$\titlefont\nwarrow$}%
\cr}}
This picture is reminiscent of \scattpic, and with good
reason\ts\rDc: projecting down from $\R^r$ onto the two-dimensional
eigenspace of $w$ for the eigenvalue $\exp(2\pi i/h)$, the
relative angles become exactly the fusing angles, and the line-lengths
exactly the masses, for a fusing of two particles of types $a$ and $b$
to form a bound state of type $\bar c$, if $c$ is the label for
the orbit containing $\g$ (so, for example, $U^c_{ab}=\pi
u(\phi_a,\b)/h$). Thus, $C^{abc}\neq 0$
should be deduced in any situation where $\b\in\G_b^+$, $\g\in\G_c$
can be found such that $\phi_a+\b+\g=0$. The picture emerging is that
each particle type should be associated with an orbit of the Coxeter
element, and it turns out that the antiparticle is associated with the
negative orbit: $\G_{\bar a}=-\G_a=\wB\G_a=\wW\G_a$. These three
different ways of conjugating the charge allows an element $\tilde
w=-\wB$ to be defined, which leaves whole orbits unchanged while
mixing around their elements in such a way that when it acts on the
three roots in \rootpic\ simultaneously, the orientation of the
projected `momentum picture', \scattpic, is reversed -- it implements
parity. But also, it is possible to show (see \rDd) that {\it all}
triplets $(\a\in\G_a,\b\in\G_b,\g\in\G_c)$ of roots satisfying
$\a+\b+\g=0$ are conjugate to each other under the combined action of
$w$ and $\tilde w$ -- there are in fact $2h$ of them, of which
$(\phi_a,\b\in\G_b^+,\g\in\G_c)$ was but one example. Hence, and by a
somewhat tortuous route, to the fusing rule for the non-vanishing of
three-point couplings in the $ADE$-related purely elastic scattering
theories:
\def\boxit#1{\vbox{\hrule\hbox{\vrule\kern4pt
           \vbox{\kern4pt#1\kern4pt}\kern4pt\vrule}\hrule}}
\setbox4=\vbox{\hbox
{~$C^{abc}\neq 0$ ~~iff $\exists$ roots
$\alpha\in\G_a, \beta\in\G_b, \gamma\in\G_c$~~with~~
$\alpha+\beta+\gamma=0$,~}
\hbox{~\phantom{$C^{abc}\neq 0$}\llap{$ie$} ~~iff~~$0\in
\G_a+\G_b+\G_c$.}}
\eqn\rule{\vcenter{\boxit{\box4}}}
Given this rule, it isn't too hard to prove that \Sabdef\ satisfies
\bootsm, and by considering projections of the root
triangles onto the other eigenspaces of $w$, consistent solutions to
the bootstrap equations for the conserved charges\ts\rZz\ can be
constructed for spins equal, modulo the Coxeter number, to an
exponent; for more details of all this, and also of how to establish the
various other properties expected of \Sabdef, see
refs.\thinspace\refs{\rDc,\rDd,\rFOa}.

\nwsec{Other examples, and twisted Coxeter elements}
This section gives very brief mention to two other situations where very
similar machinery is encountered. The first of these is a calculation
by Saleur and Bauer\ts\RF\rSBa\SBa\ of the partition functions
of the Pasquier models\ts\RF\rPc\Pc\ on a cylinder. They found that if the
heights were constrained to be equal to $a$ and $b$ (two nodes on the
relevant Dynkin diagram) at the two ends of the cylinder, then in the
continuum limit the partition function could be expanded in Virasoro
characters as
\eqn\Zijdef{Z_{ab}(l,l')
\sim\sum_{\l=1}^{h{-}1}V^{\l}_{ab}\chi_{1,\l}(q),}
where
\eqn\vdef{V^{\l}_{ab}=\sum_{s\in\{{{\rm exponents}\atop{\rm of~}g}\}}
{\sin(\pi s\lambda/h)\over\sin(\pi s/h)}q^{(s)}_aq^{(s)}_b.}
Here,
$\chi_{1,\l}$ is a Virasoro character from the first row of the Kac
table for the central charge $c=1-6{/}h(h-1)$ of the $g$ model,
$q=\exp(-\pi l/l')$ is the modular parameter for a cylinder of
circumference $l$ and width $l'$, and $q^{(s)}$ is an eigenvector of
the Cartan matrix of $g$, with eigenvalue $2{-}2\cos \pi s/h$.
Subsequently, the $V^{\l}_{ab}$ were also studied in the context of
general models based on graphs, and fusion
algebras\thinspace\NRF\rDZa\DZa\NRF\rSg\Sg\refs{\rDZa,\rSg}. The
connection with the material of section 2 comes from the observation
that the sum \vdef, if non-zero, simply gives the expansion
of the inner product
between a root and a weight of $g$, in a basis of eigenplanes of the
Coxeter element\ts\rDd. More precisely,
\eqn\obs{V^{1+u(\phi_a,\b)}_{ab}=(\l_a,\b)\,,}
and \Zijdef\ can be rewritten in a way very reminiscent of \Sabdef:
\eqn\Zijdefii{Z^{(G)}_{ab}
\sim\sum_{\b\in\G^+_b}(\l_a,\b)\chi_{1,1{+}u(\phi_a,\b)}\,.}
One consequence is that the tables \tableone\ and \tabletwo\ can also
be thought of as lists of partition functions. More important is that
the positivity of the root $\b$ in \Zijdefii\ establishes the
positivity of the $V^{\lambda}_{ab}$'s, expected from their appearance
in \Zijdef\ as multiplicities, in a general way. Previously this had
only been checked case-by-case.

The S-matrices for the (simply-laced) affine Toda field
theories\thinspace\NRF\rAFZa\AFZa\NRF\rCMe\CMe%
\refs{\rAFZa,\rCMe,\rBCDSb,\rBCDSc}\ are
rather more obviously related to the earlier discussion. Essentially,
the only problem with \Sabdef\ in this context is its lack of
a coupling-constant
dependence, and this is easily remedied by replacing the block
$\ubl{x}$ defined in \ubldef\ by a slightly more complicated object,
namely
\eqn\Bbl{\ubl{x}_B=
 {\usbl{x-1}\usbl{x+1}\over \usbl{x-1+B}\usbl{x+1-B}}\,,}
where $B$ contains the coupling constant:
$B(\b)=2\b^2/(\b^2{+}4\pi)$. After this modification, all the earlier
discussion of physical-strip pole structure, fusing rules and so on
goes through unchanged. However an affine Toda theory also
has a Lagrangian, which can be expanded
perturbatively to find the classical three-point couplings. Case-by-case,
their non-vanishing was known to be the same as that deduced from the
quantum S-matrices; but more recently, a general group-theoretic proof
that the classical couplings obey the rule \rule\ has been
given\thinspace\RF\rFb{\Fb\semi\FLOa}. The connection between
\rule\ and a previously-observed Clebsch-Gordon selection rule has
also been established\ts\RF\rBa\Ba.

The quantum theory of the non simply-laced affine Toda models is
considerably more complicated than that of the simply-laced
versions\ts\rDGZa, and to find a geometrical understanding remains a
challenge. However, in the classical domain, the conserved
charge bootstrap, and the treatment\ts\rFb\ of the three-point
couplings, go through unharmed,
providing a uniform description of all the untwisted affine theories,
based on the Coxeter elements of the
underlying (non-affine) Weyl groups. One gap has been the twisted non
simply-laced cases, and it seems worth pointing
out that in fact the necessary concept has already been introduced by
Steinberg, and is described in
an article by Springer\ts\RF\rSj\Sj. It is the `twisted Coxeter
element', defined as follows. Recall first that whenever a Dynkin
diagram has an automorphism, $\sigma$ say, then the automorphism group
of the root lattice $\Phi$ is larger than $W$, the Weyl group -- there
are also `outer automorphisms', induced by $\sigma$.
The twisted Coxeter element for $\sigma$ lies in $W\sigma$, and is
defined by first choosing one simple reflection from each
$\sigma$-orbit in the simple roots (note, $\sigma$, being a diagram
automorphism, maps the simple roots to themselves) and then forming
the product $w'$ of these reflections. Then
$w_{\sigma}\equiv w'\sigma$ is a twisted Coxeter element, a particular
outer automorphism of the non-affine root system $\Phi$. Combining the
information in \rSj\ with a remark in an article by
Kac\thinspace\RF\rKh{\Kh, remark on p.\ts 127}
is enough to see that the arguments of \rFb\ will
go through in this case as well, leading to a characterisation of the
twisted affine couplings in terms of the orbits of the twisted Coxeter
element. The properties of $w_{\sigma}$ are sufficiently simple (for
example, all orbits in $\Phi$ have the same length)
that it is tempting to try to generalise \Sabdef\ to
this case. One problem is that the orbits are longer, and hence there
are fewer of them -- certainly less than $r$, the dimension of the
space in which $\Phi$ sits. Thus, if a particle type is associated
to each orbit, the S-matrix formula \Sabdef\ cannot be used as it
stands since the term $(\l_a,\b)$ in the exponent links particle $a$
to a fundamental weight, of which there are now too many. The relation
\wtrel\ provides a hint as to one way out, suggesting  the
expression
\eqn\Sabtdef{S_{ab}=\prod_{\b\in\tilde\G^+_b}
\{1+u(\tilde\phi_a,\b)\}^{((1-w_{\sigma}^{-1})^{-1}\tilde\phi_a,\b)}}
where $\tilde\G_b^+$ is now the positive part of a $w_{\sigma}$-orbit,
and $\tilde\phi_a$ is the (unique for suitably-chosen simple roots\ts\rSj)
positive root in $\tilde\G_a$
which becomes negative on the action of $w_{\sigma}$. Unfortunately,
although \Sabtdef\ produces functions which satisfy all the
bootstrap requirements in terms of bound-state structure and bootstrap
consistency, it cannot be the right answer for the quantum theory. In
fact, it yields sub-matrices of the simply-laced
S-matrices. To see why this should be so, an alternative
characterisation of the twisted Coxeter element can be used. Recall from
\RF\rOTa\OTa\ that each twisted affine Dynkin diagram can be obtained as a
folding of some simply-laced affine diagram, a fact
useful in establishing the classical properties of the twisted
affine Toda theories. For example, the $d^{(3)}_4$ diagram can be
obtained by folding that for $e^{(1)}_6$, so that the solutions of the
$d^{(3)}_4$ affine Toda theory can be found simply by imposing a
certain symmetry on the solutions of the $e^{(1)}_6$ model -- a
process known as reduction. Now let $w$ be a Coxeter element for the
non-affine root system $E_6$, and let $P$ be the
$w$-eigenplane for the exponent $4$ of $E_6$, that is for the eigenvalue
$\exp{2\pi i/3}$; it is also the eigenplane for the exponent
$8$. The orthogonal complement of this plane, $P^{\bot}$, is
four-dimensional, and its intersection with $\Phi$, the set of $E_6$
roots, is a set of roots for $D_4$, made up from the $w$-orbits of the
two self-conjugate particles in the $e_6^{(1)}$ affine Toda model.
Furthermore,
the action of $w$ in this embedded $D_4$ is precisely that of a $D_4$
order-three twisted Coxeter element. An
inner automorphism of $E_6$ has induced an outer
automorphism of $D_4$, a fact which explains why the three point
couplings in the twisted folding were observed to be simply a subset
of those of the parent theory\thinspace\rBCDSc, and also explains why
the S-matrix predicted by \Sabtdef\ for this case is just that for the
scattering of the two self-conjugate particles in the $E_6$ theory.
Although I do not know a general proof, a case-by-case check shows
that this phenomenon generalises to all the twisted foldings.

While \Sabtdef\ does not reproduce the formulae of \rDGZa\ for the
twisted affine Toda theories, it only fails at the last
hurdle, not accounting correctly for some quantum
effects. In this respect the situation here is in
better shape than for the untwisted non simply-laced models, where
the simplest guess would be just to apply formula \Sabdef\ with $w$ a
Coxeter element of the relevant non simply-laced algebra. Since the
roots and weights are then no longer dual, the exponents cease to be
integers and the analyticity properties of $S_{ab}(\t)$
are drastically changed. In fact, for the non simply-laced theories,
the quantum duality
between strong and weak couplings is expected to relate
the untwisted and twisted
theories\ts\NRF\rWWa\WWa\NRF\rKWb\KWb\refs{\rDGZa,\rWWa,\rKWb}.
It seems that the rigid structure organising the simply-laced cases is
being deformed into something rather richer, and a deeper
understanding of this would be very interesting.

\medskip\noindent{\titlerms Acknowledgements}\smallskip\nobreak
I would like to thank the Isaac Newton Institute for its
hospitality during the writing of this paper, and Ed Corrigan and
Gerard Watts for interesting discussions.
The work was supported by a grant under the
EC Science Programme.

\listrefs
\bye